\documentclass[useAMS,fleqn, usenatbib, final]{mnras}
\usepackage{amsmath, amssymb, mathtools}
\usepackage{graphicx}
\usepackage{dcolumn}
\usepackage{color,units}
\usepackage[dvipsnames]{xcolor}
\usepackage{lineno}
\usepackage[title]{appendix}
\usepackage{xspace}
\usepackage{caption}
\usepackage{longtable}
\usepackage{makecell}
\usepackage{aas_macros}

\usepackage{hyperref}
\usepackage{mwe,subcaption, tikz}
\tikzset{boximg/.style={remember picture,red,thick,draw,inner sep=0pt,outer sep=0pt}}

\newcommand{\bilby}{\texttt{bilby} }
\newcommand{\tensorflow}{\texttt{TensorFlow} }
\newcommand{\gpflow}{\texttt{GPFlow} }

\newcommand{\Cardiff}{Cardiff University, Cardiff CF24 3AA, UK}

\begin{document}

\title{Density estimation with Gaussian processes for gravitational-wave posteriors}
\author[d'Emilio et al.]{
\parbox{\textwidth}{
V.~D'Emilio, 
R.~Green, 
V.~Raymond, 
}\vspace{0.2cm}\\
\Cardiff\\
}


\date{\today}

\maketitle

\begin{abstract}
The properties of black-hole and neutron-star binaries are extracted from gravitational-wave signals using Bayesian inference. This involves evaluating a multi-dimensional posterior probability function with stochastic sampling. The marginal probability density distributions from which the samples are drawn are usually interpolated with kernel density estimators. Since most post-processing analysis within the field is based on these parameter estimation products, interpolation accuracy of the marginals is essential. In this work, we propose a new method combining histograms and Gaussian Processes as an alternative technique to fit arbitrary combinations of samples from the source parameters. This method comes with several advantages such as flexible interpolation of non-Gaussian correlations, Bayesian estimate of uncertainty, and efficient re-sampling with Hamiltonian Monte Carlo.

\end{abstract}

\section{Introduction}\label{sec:intro}
The first detection of gravitational waves (GW) in 2015 \citep{abbott2016observation} sparked a new era of Astronomy. Several years on from that event the number of detected gravitational waves keeps increasing and within this decade we expect to observe $O(10^3)$ signals \citep{abbott2020prospects} from compact binaries coalescences (CBCs). This huge progress brings with it the challenge of efficiently analysing a large number of events. To address these computational challenges, machine learning techniques are being increasingly investigated within the field of gravitational-wave physics \citep{cuoco2020enhancing}. 
Many studies have focused on speeding parameter estimation of the source parameters of the signals with various techniques, such as deep learning \citep{george18}, variational autoencoders \citep{gabbard19} and autoregressive neural flows  \citep{green2020gravitational}. Other work has focused on combining detection and parameter estimation with deep neural networks \citep{fan2019applying} as well as using neural networks to rapidly generate surrogate waveforms \citep{khan2020gravitational, chua2019reduced}.

While the research efforts to speed up or completely revolutionise parameter estimation are ongoing, the issue of how to effectively deal with a large number of results from different events remains. In particular, how to streamline the analysis of the results, while maintaining accuracy.
In this work, we demonstrate the efficiency and usefulness of using Gaussian processes (GP) for post-processing parameter estimation results of CBCs. 
Applications of GPs in the field of gravitational waves span a wide range of use-cases, such as marginalising waveform errors \citep{Moore2016}, regression of analytical waveforms \citep{setyawati2020regression}, predictions of population synthesis simulations \citep{Barrett2016}, hierarchical population inference \citep{Taylor2018} and Equation of State (EOS) calculations \citep{landry2019nonparametric}.
They have also been exploited for the development of fast parameter estimation with \texttt{RIFT} sampler \citep{lange2018rapid}.

Here we exploit GPs to estimate probability density functions (PDFs) from parameter estimation of gravitational-wave signals. Non-parametric density estimation from a finite set of samples is an active research field in machine learning and statistics \citep{ murray2008gaussian, papamakarios2017masked, wang2019nonparametric}.

For most GW analysis, histograms are usually the preferred estimators to visualise the marginal posterior PDFs and to avoid over-smoothing sharp features, but often are not convenient for post-processing analyses such as population inference. These sorts of analyses either re-weight the posterior samples directly \citep{abbott2020population} or need to estimate a continuous representation of the gravitational-wave posterior density surface. Several density estimation methods such as Dirichlet process' \citep{del2018dirichlet}, Gaussian Mixture Models \citep{talbot2020fast} and others have been employed to address this problem specifically for gravitational-waves. As well as these, A closely related method to GPs \citep{kanagawa2018gaussian}, Gaussian Kernel Density Estimators (KDEs) are sometimes employed in gravitational waves' analyses \citep{pitkin2018hierarchical, pang2020parameter, lynch2017information}.

These KDEs are often effective but they assume correlations between parameters to be linear and smooth, making this method sometimes limited in flexibility. There exist many variations of the KDE algorithm to take into account specific interpolations problems, but there isn't a single implementation that is guaranteed to be robust against all possibilities. A specific KDE implementation might solve an issue in one case and be the cause of some inaccuracies in another~\cite{wand1994kernel}. 

We implement a single technique that can interpolate arbitrary multi-dimensional slices in parameter space, which can handle both simple and difficult space morphology, such as sharp bounds and non-Gaussian correlations. Our modelling tool is based on the histogram density estimate, combining the histogram's accurate treatment of the samples' features with the predictive capabilities of GPs.
An additional advantage of this technique is that it can provide a Bayesian measure of uncertainty from the finite (and sometimes small) number of samples for post-processing analysis. This measure of model uncertainty could then be incorporated into any analysis where the marginalised posterior density is used.

In Sec.~\ref{sec:methods} we describe our density estimation technique in the context of gravitational-wave parameter estimation and machine learning. We propose a series of example applications in Sec.~\ref{sec:results}, which allows us to discuss the advantageous features of our method. Finally, in Sec.~\ref{sec:conclusions} we summarise our findings and suggest future extensions of this work.

\section{Methods}\label{sec:methods}
In this section, we introduce the mathematical framework of the techniques discussed. In subsection~\ref{sec:pe-intro} we discuss the Bayesian inference problem for gravitational waves and the density estimation techniques currently employed in the field. In subsection~\ref{sec:density_est_with_gps}, we outline the fundamentals of GPs and their interpretation for interpolating a posterior density surface. We then describe the details of our GP implementation and how to model probability densities from parameter estimation. 
\subsection{Bayesian inference and density estimators}\label{sec:pe-intro}
We describe the gravitational-wave data in the detector $d$ as the sum of a waveform model $h(\vec{\theta})$ and a combination of instrumental noise, which we assume to be Gaussian. The probability of observing data parametrised by $\vec{\theta} = \{\theta_{1},\theta_{2},...,\theta_{N}\}$, can be defined as the probability that  $r = d - h$ is the realisation of the instrumental noise.
This likelihood can be written as :

\begin{equation} \label{likelihood_calculation}
        p(d|\vec{\theta}) \propto \exp\left(-\frac{1}{2}\langle d-h(\vec{\theta}) | d - h(\vec{\theta}) \rangle \right),
\end{equation}
where $\langle a | b \rangle$ denotes the inner product between two waveforms $a$ and $b$ and is defined as \citep{cutler1994gravitational},
\begin{equation}\label{eq:inner_prod}
        \langle a | b \rangle = 4\text{Re}\int_{0}^{\infty}{ \frac{\tilde{a}(f)\tilde{b}^{*}(f)}{S_{n}(f)} df},
\end{equation}
where $S_{n}(f)$ is the one-sided power-spectral density (PSD) and $\tilde{a}$ denotes the Fourier transform of the gravitational waveform $a$.
\par

By choosing astrophysically motivated priors over the model parameters, we can use the Bayesian framework to calculate the posterior probability distribution for the source parameters \citep{thrane2019introduction}

\begin{equation}
  \begin{aligned}
p(\vec{\theta} | d ) &= \dfrac{(p(d| \vec{\theta} )p(\vec{\theta})}{p(d)}    \\
&\propto p(d | \vec{\theta} )p(\vec{\theta})
  \end{aligned}
  \label{eq:log-joint}
\end{equation}

The posterior probability is generally a $15$-dimensional surface for a circular binary black hole (BBH) merger but can be $17$-dimensional in the case of a binary neutron star (BNS) merger. The dimensionality depends on the physical parameters describing the signals. Generally, these are distinguished between extrinsic parameters, such as sky localisation, and intrinsic parameters, such as the masses of the sources.

Posterior distributions for specific parameters can then be found by marginalising over all other parameters,
\begin{equation} \label{posteriorforoneparameter}
        p(\theta_{i}|d) = \int{p(\vec{\theta}|d) d\theta_{1}...d\theta_{i-1} d\theta_{i+1}... d\theta_{N}}.
\end{equation}
The posterior is however generally intractable and therefore must be evaluated via stochastic methods such as Markov Chain Monte Carlo (MCMC) and nested sampling, these are implemented (and specifically tuned for the gravitational wave problem)  in Bayesian inference packages such as LALInference \citep{veitch2015parameter} and \bilby \citep{ashton2019bilby}.

\subsection{Density estimation with Gaussian Processes } \label{sec:density_est_with_gps}
\subsubsection{\label{sec:definition-gp} Definition and interpretation}
GPs are interpolation methods with a probabilistic interpretation, they are built on a Bayesian philosophy, which allows you to update your beliefs based on new observations. 
The process can be understood as an infinite distribution over functions, such that any finite collection of these has a Gaussian distribution. As data is observed, the GP is \textit{conditioned} and the range of possible functions that can explain the observations is constrained. 
As such a GP is defined by 
a mean, which represents the expectation value for the best fitting function, and by a covariance matrix, called a kernel, which measures the correlations between observations ~\citep{williams2006gaussian}. 
In the absence of observations, the GP predictions will revert to a prior mean function, which is usually chosen to be zero, and which properties are determined by the kernel architecture.
Mathematically this is written as:

\begin{equation}
    f(\vec{x}) \sim \mathcal{GP}(m(\vec{x}), \kappa(\vec{x}, \vec{x}{\prime}))
\end{equation}
where the mean and covariance are denoted as:

\begin{align*}
    \mu(\vec{x}) &= \mathbb{E}\left[f(\vec{x})\right]\\
    \kappa(\vec{x}, \vec{x}^{\prime}) &= \mathbb{E}\left[(f(\vec{x}) - m(\vec{x}))(f(\vec{x}^{\prime}) - m(\vec{x}^{\prime}))\right]
\end{align*}

We can then model a surface y conditioned on our observations as:

\begin{equation}
    y_* |f, x \sim \mathcal{N}(f(x_*),  \sigma^2_*)
\end{equation}
where, in this application, the dimensionality of $f$ will depend on how many parameters $p(\theta_{i}|d)$ has been marginalised over.

The non-parametric nature of GPs makes this technique flexible, but it can be computationally expensive as the whole training set needs to be taken into account at each prediction. The standard implementation has $\mathcal{O}(N^3)$ computations and $\mathcal{O}(N^2)$ storage, this then becomes prohibitive for $\sim10 \textrm{k}$ data observations or more.
To tackle this issue it is common to use sparse inference methods, which approximate the conditioning of the GP over a set of $ M << N $ `inducing' points.
The evaluation over the inducing points $M$ is then much cheaper than for an `exact' GP resulting in $\mathcal{O}(NM^2)$ computations rather than $\mathcal{O}(N^3)$ ~\citep{quinonero2005unifying, hensman2013gaussian}. As well as sparse methods one can exploit multi-GPU parallelization and methods like linear conjugate gradients to distribute the kernel matrix evaluations which then allows for exact inference to be performed on a short time scale~\citep{wang2019exact}. In this work, however, we find that sparse approximations were accurate enough to effectively model the marginalised posterior surfaces that we were interested in. Moreover, once a GP has been `trained' over the data, it is possible to draw infinitely many function realisations from it without recomputing the expensive covariance matrix.

A recognised advantage of GPs is reliable uncertainty estimate when making predictions over unseen data. In this application, we are not interested in predicting the value of the posterior in unexplored regions of the parameter space, but only in generating a faithful model where we have posterior samples. In regions within the space of parameters, the GP variance depends on our choice of training points, which is useful to assess the accuracy of our density estimation. In terms of uncertainty estimation this can be explained as our model having very low \emph{epistemic} uncertainty everywhere, we then seek to estimate the \emph{aleatoric} uncertainty due to our model fit around the random fluctuations in the histogram densities which are used to train the GP. 

\subsubsection{\label{sec:model-gp} Model construction}
In this application, we want to use a GP to estimate the marginalised posterior density for any subset of parameters. We train our GP using the normalised histogram counts over a grid of points, i.e. the centroids of the histogram bins, that cover the marginalised parameter space.  We then fit our GP to this discrete set of points to generate a continuous representation of the surface.

We employ \tensorflow and \gpflow to implement our GP modeling infrastructure, which includes two inference schemes: exact inference for 1-2 dimensional problems ($\mathcal{O}\sim1000$ samples) and sparse inference for higher dimensionality due to computational costs.  As well as a difference in the inference scheme, when extending this method to higher dimensions, our choice of training data changes. When creating the grid over four dimensions we find that the typical set has a volume of $O(1\%)$,(this is a common problem associated with the curse of dimensionality). We, therefore, choose to discard the empty bins and encode our knowledge of these points through the choice of prior over our GP.

Since the model is constructed with converged posterior samples, there is no probability support where the histogram bins are empty. To encode this, we set the mean of the GP to be equal to zero, such that far away from the training data the model will have a high variance but a mean of zero.

To estimate the density for a given region of parameter space we then simply evaluate the GP at those parameters, i.e. 
\begin{equation} \label{eq:gp_density_approx}
    \begin{array}{l}
        p(\vec{\theta} = \vec{x_*} |d) \approx y_* |f, x \\
        \quad \quad \quad \quad \quad  \sim \mathcal{N}(f(\vec{x_*}),  \sigma^2_*)
    \end{array}
\end{equation}

Due to bounded priors (e.g at mass ratio $m_{2}/m_{1}\coloneqq q=1$), the posterior surface often presents sharp discontinuities and therefore the surface is only \emph{piece-wise continuous}. GPs are in principle flexible enough to model any surface including piece-wise continuous ones, however, we found in practice that it is more favourable to decompose our density function into two components, one smooth, continuous function, and one step function. We do this by multiplying the density and our GP estimate by a step function, which is zero at any discontinuities and 1 otherwise.

\begin{align*}
\pi(\vec{x_{*}}) = 
\begin{cases}
1 & \text{if  $x_{min} < \vec{x_{*}} < x_{max}$}\\
0 & \text{otherwise}
\end{cases}
\end{align*}

Multiplying by this step function is then analogous to imposing a prior over our posterior surface, i.e. it allows us to rewrite the equation \ref{eq:gp_density_approx} as 

\begin{equation}
    \begin{array}{l}
        p(\vec{\theta} = \vec{x_*} |d)  \pi(\vec{x_*}) \approx (y_* |f, x) \pi(\vec{x_*})\\
        p(\vec{\theta} = \vec{x_*} |d) \quad  \quad \sim \mathcal{N}(f(\vec{x_*}),  \sigma^2_*) \pi(\vec{x_*})
    \end{array}
\end{equation}

We are free to encode our knowledge in this way and perform the decomposition as we do not change the original posterior surface that we would like to model in any way. This enhances the robustness of the model against all discontinuities, including artificial cuts in parameter space that might be required for post-processing analysis.

The variance of the GP depends on the kernel, but also on the noise variance parameter of the likelihood. Usually, the noise variance is given by a single number, i.e. homoskedastic noise, which reflects the random fluctuations of the posterior samples. In low-dimensional examples, where we employ an exact inference scheme, we can assign multiple values to the noise variance, i.e. heteroskedastic noise~\citep{mchutchon2011gaussian}. In such instances, we are then able to propagate the error from the histogram on the density estimate, which is simply given by the Poisson noise in each bin $\sigma_{\textrm{bin}}\sim\sqrt{N_{\textrm{counts}}}$. Incorporating heteroskedastic errors within a sparse inference scheme is an area of current research in the field of machine learning~\citep{liu2020large}. 

It is common practice to build an interpolation of a posterior surface in order to draw more samples from it.
As our model is implemented in \tensorflow we can quickly draw more samples from the marginalised posteriors using the many samplers available in the package library, such as Hamiltonian Monte Carlo (HMC)~\citep{betancourt2017conceptual}.
Further technical details which had a significant impact on our model accuracy, such as our data pre-processing scheme and our choice of kernel design are included in Appendix~\ref{apdx:techincal-gp}.

\section{Results}\label{sec:results}
In this section, we present our model and a series of example applications for gravitational waves. In Sec.~\ref{sec:GP-analytical} we illustrate the method on a simple 1-dimensional analytical example. In Sec.~\ref{sec:GP-examples} we show examples of common post-processing applications for our density estimation tool. Finally, we discuss our treatment of GP model uncertainty and how we propagate it to produce uncertainty on the marginalised posterior distributions.
\subsection{ Analytical 1D example}\label{sec:GP-analytical}
Our proposed GP modelling technique is by construction flexible and robust against all distribution morphologies.
To illustrate this, we construct a non-trivial 1-dimensional example: an inverse gamma function $f(x, \alpha) = \frac{x^{-\alpha-1}}{\Gamma(a)} \exp(-\frac{1}{x})$, with  $\alpha=2$ and a sharp bound at $x = 0.75$. 

\begin{figure}
    \centering
    \includegraphics[width=1.05\linewidth]{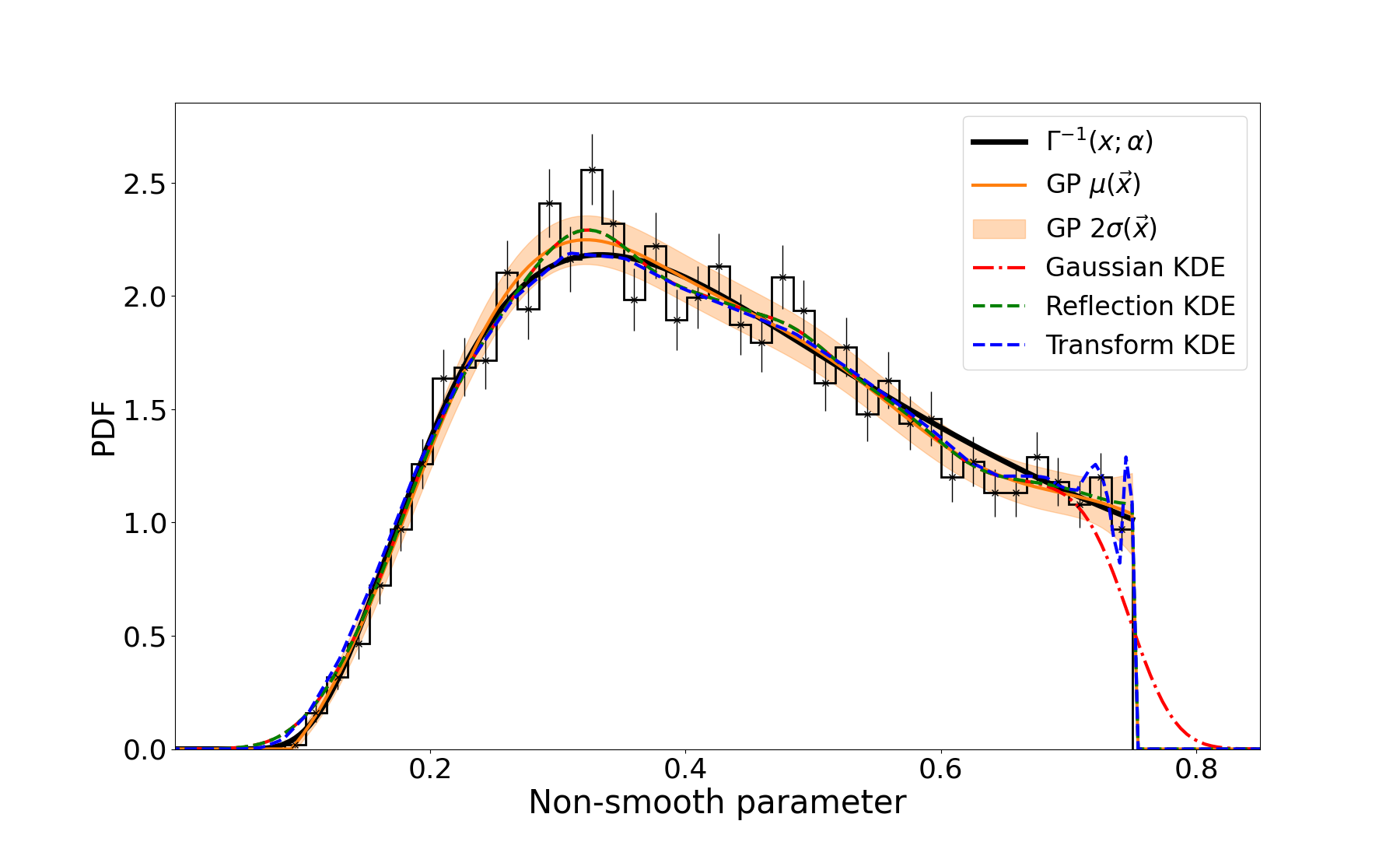}
    \caption{Interpolation of a bounded one-dimensional inverse gamma density function (in solid black) with our GP-based method (in solid orange). The histogram points used to generate the model and its uncertainty are shown as black points with error bars. Alternative KDE methods are shown for comparison as coloured dashed lines.}
   \label{fig:1D_KDE}
\end{figure}

In Figure~\ref{fig:1D_KDE} we show our GP model mean prediction and uncertainty, compared to a Gaussian KDE from \texttt{scipy.stats}~\cite{virtanen2020scipy} and two KDE transformations implemented in \texttt{PESummary}~\citep{hoy2020pesummary}, a commonly used post-processing package in gravitational-wave astronomy. The \textit{reflection} and \textit{transform} KDEs, are examples of augmentations on the standard (Gaussian) KDE, and are generally used to model difficult features introduced at the boundaries of posterior distributions. Both of these improvements to the standard KDE apply a transformation at the boundary which implicitly assumes some distributional features (see ~\citep{hoy2020pesummary} for more details). A Gaussian Process on the other hand makes no assumptions about the distributional shape and can in principle fit any distribution. 

We show an example in Fig \ref{fig:1D_KDE} where our GP is able to well model the posterior and the \textit{reflection} KDE provides a better fit than the other KDE methods. The \textit{transform} KDE is more sensitive to noisy features in the samples and can present artifacts, while the Gaussian KDE over-smooths the sharp cut at $0.75$.  Following this illustrative example, there are others where the \textit{reflection} KDE is less appropriate. This example was chosen to highlight a case where the choice of KDE is important to fit the distribution well. While synthetic and not representative, it does illustrate features that can and do happen in gravitational-wave astronomy when analysing posteriors. In examples such as this our GP model provides an alternative method to KDEs, requires less hand-tuning, and also provides a Bayesian estimate of the error on the density estimate, as propagated from the histogram errors.



\subsection{\label{sec:GP-examples} GW Applications}
We now look at a few important post-processing problems in gravitational-wave astrophysics. The training time required to generate the models presented in this section is of the order $\mathcal{O}(2mins)$, with variations due to the dimensionality of the surface and to the inference scheme employed. 
To assess the quality of the model in more than one dimension we decide to re-sample the surrogate surface and compare the new samples to the original set, part of which has been used for training.
All samples used in the following sections are taken from the Bilby GWTC-1 catalog~\citep{romero2020bayesian}. 

\subsubsection{\label{sec:GP-intrinsic} Catalogue of gravitational-wave properties }
Gravitational-wave detection parameters can be distinguished between those intrinsic to the sources, such as the component masses, and those extrinsic to them, such as the sky location. Interpolating the marginal posteriors of combinations of these parameters is often necessary for post-processing. The following example illustrates a simple case where one can use a GP to interpolate the intrinsic parameters for a given detection. In practice, this could then be repeated for entire GW catalogues so that these interpolated posterior surfaces are then combined for population inferences on the sources of GWs.

\begin{figure}
    \centering
    \includegraphics[scale=0.3]{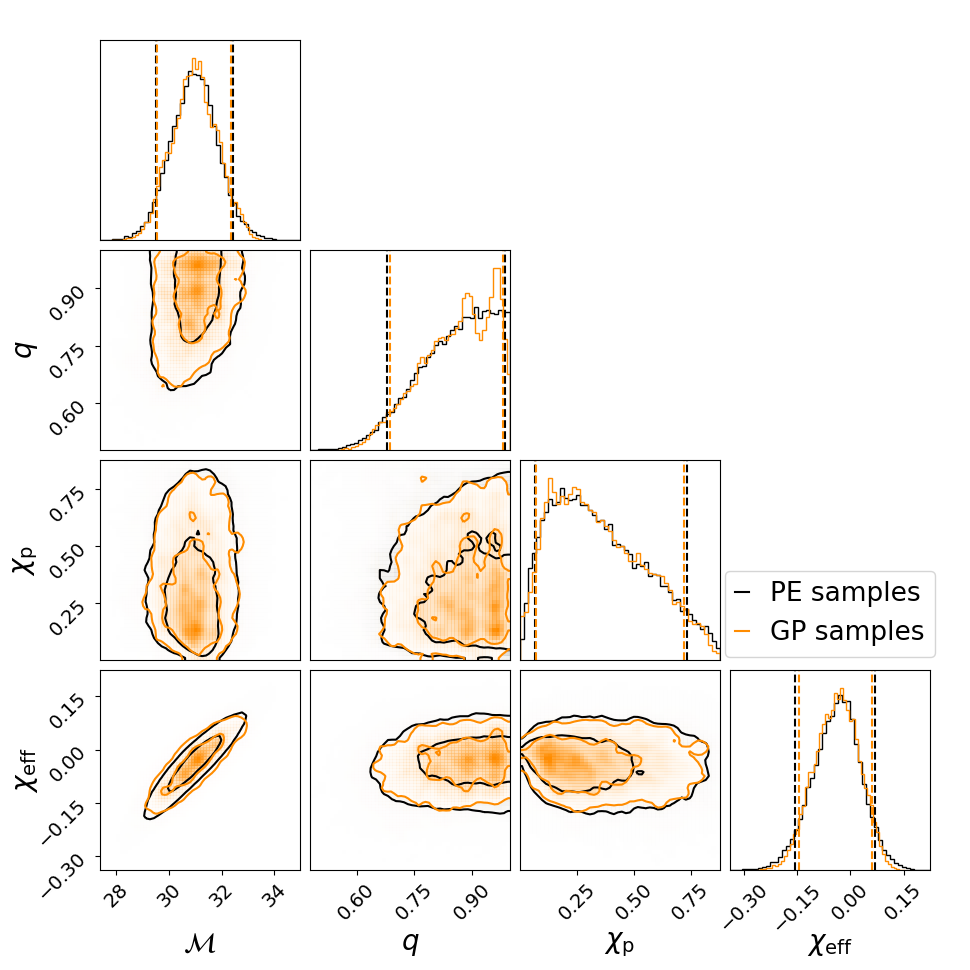}
    \caption{Corner plot of the intrinsic parameters of GW150914, drawn from our GP surrogate (in orange) compared to the original PE samples (in black).}
    \label{fig:GP_intrinsic}
\end{figure}

For this example, we interpolate the marginal posterior distribution of the intrinsic parameters of the first BBH detection GW150914~\citep{abbott2016properties}, parametrised as follows: chirp mass $\mathcal{M}$, mass ratio $q= m_{2}/m_{1}$ (where $m_{1}>m_{2}$), effective inspiral spin component $\chi_{\textrm{eff}}$ and effective precession spin $\vec{\chi_{\textrm{p}}}$, defined by the spin components that lie in the orbital plane~\citep{schmidt2015towards}.
In Figure~\ref{fig:GP_intrinsic} we compare the marginal distributions sampled from our GP model to the original PE samples. We can visually assess that the correlations between parameters are accurately reconstructed as the 50\% and 90\% contour lines overlap for each pair of parameters. In Table~\ref{tab:intrinsic_CI} we report the mean and 90\% confidence intervals of the samples drawn from our model and which we find in agreement to the values from the original samples within the expected uncertainty.

\begin{table}
    \centering
    \begin{tabular}{c|c|c}
& \thead{GP samples} & \thead{PE samples}\\
\hline
\makecell{Chirp mass $\mathcal{M}/M_{\odot}$}& ${30.95}_{-0.97}^{+0.93}$ & ${30.96}_{-0.89}^{+0.86}$\\
\makecell{Mass ratio $q$}& ${0.87}_{-0.12}^{+0.09}$ & ${0.87}_{-0.12}^{+0.09}$\\
\makecell{Effective precession \\ spin component $\chi_{\rm{p}}$} & ${0.33}_{-0.19}^{+0.26}$&
${0.32}_{-0.19}^{+0.27}$\\
\makecell{Effective inspiral \\ spin component $\chi_{\rm{eff}}$} & ${-0.04}_{-0.07}^{+0.07}$ & ${-0.04}_{-0.07}^{+0.06}$

    \end{tabular}
    \caption{Source properties of the intrinsic parameters of GW150914, original samples and samples from the GP interpolation.}
    \label{tab:intrinsic_CI}
\end{table}
\subsubsection{\label{sec:GP-EOS} Accurate interpolation for conditional integrals}
Many astrophysical inquiries in gravitational-wave astronomy require evaluating conditional integrals across parameter space, which in turns require sampling additional posterior points constrained to a hyperplane. This is for instance the case when estimating the Equation of State (EOS) from BNS collisions, an important post-processing analysis that allows us to probe extreme conditions of matter \citep{abbott2018gw170817}. 
This is possible because the compactness of the objects is imprinted in the gravitational waveform and can be measured by the tidal deformability parameters. The EOS integral involves evaluating the marginal posterior distribution over the masses $(\mathcal{M}, \eta)$ and tidal parameters $(\Tilde{\Lambda}, \delta\Tilde{\Lambda})$, subject to constraints between those parameters as parametrised by the EOS.
 
\begin{figure}
    \centering
    \includegraphics[scale=0.3]{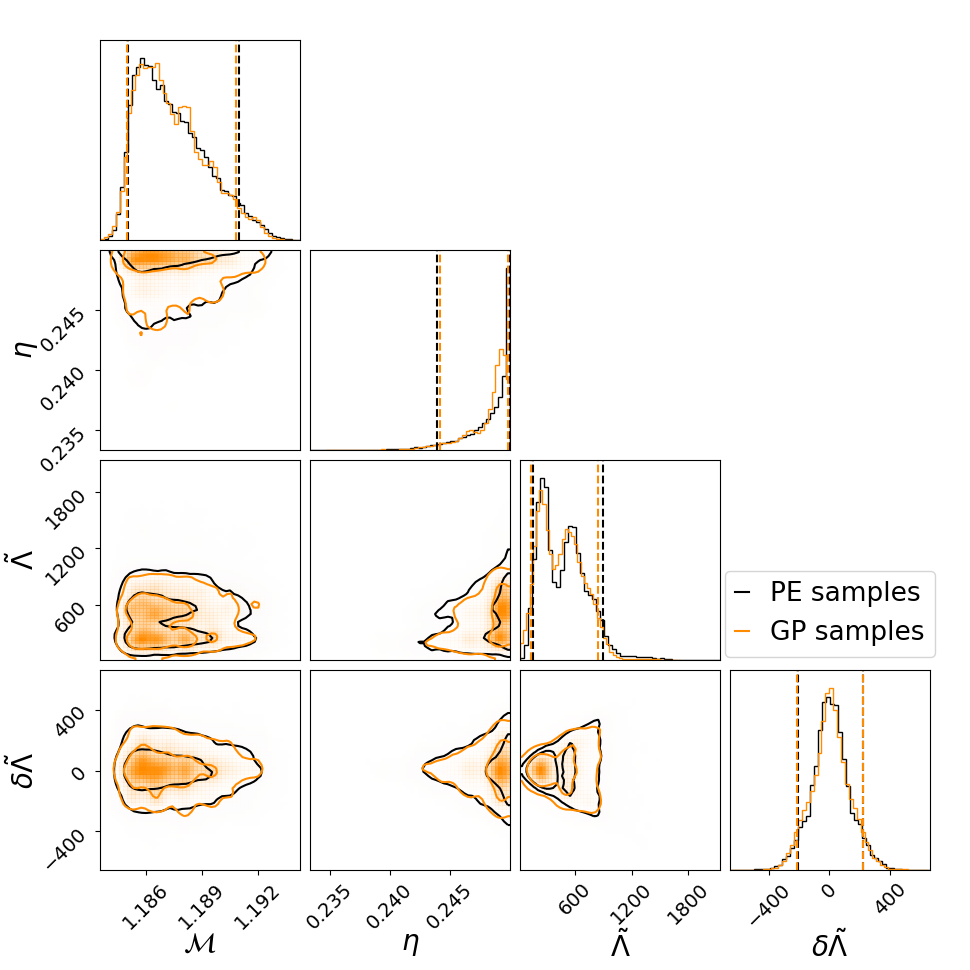}
\textbf{}    \caption{Corner plot of the mass and tidal parameters of GW170817, drawn from our GP model (in orange), compared to original PE samples in black.}
    \label{fig:GP_EOS}
\end{figure}

There are instances where the marginal posterior for these parameters contain non-linear correlations, as is the case for the first BNS event GW170817~\citep{abbott2017gw170817}.
We test our interpolation model over this 4-dimensional surface.
In Figure~\ref{fig:GP_EOS} we compare the marginal distributions sampled from our GP model to the original PE samples. We see that our GP is able to faithfully represent the marginalised posterior surface, in particular, we see that there is good agreement between the $90\% $ credible intervals. When looking at the 2d contours see that the 50\% and 90\% levels agree very well and that the GP model is able to capture degenerate features and bi-modalities. Finally, our interpolation of the surface can be re-sampled efficiently and for this example, we obtained 750k samples in
a few $\mathcal{O}(5mins)$ (depending on hardware) using an HMC sampler.
Hence this method can be advantageous over traditional methods, where the interpolation is generally performed with a Gaussian KDE by transforming the symmetric mass ratio parameter to be $log(0.25 - \eta)$ \citep{pang2020parameter} and there is no measure of uncertainty over the fit. 
\subsubsection{\label{sec:GP-sky-loc} Propagating GP uncertainty}
GPs provide a fully Bayesian estimation of the uncertainty over model predictions, as the full covariance matrix between posterior samples is computed.
In each of the GW applications shown so far we have utilised the mean prediction of the GP function. This uncertainty measurement can be very important in many cases, however here we illustrate with a single example how one can extract the uncertainty from the modelling.
Accurate localisation of a gravitational signal can be of fundamental importance for multi-messenger astronomy~\citep{grover2014comparison, ligo2017multi} and for measurements of cosmological parameters with dark sirens~\citep{soares2019first}. As the localisation accuracy decreases, the marginal posteriors for the sky location parameters can look degenerate and non-Gaussian. 
We build an interpolation of the sky location parameters, right ascension ($ra$) and declination ($dec$), of GW150914. This event was observed by only two detectors, so albeit its high SNR, its sky location presents a typical ring-like shape.

The uncertainty measure produced by the GP is a Gaussian distribution about any given point on the surface, when considering the entire surface the combination of these Gaussians can be interpreted as a range of plausible density surfaces for any given confidence level (e.g. $2\sigma$). The uncertainty on the 1D marginal distributions can then be calculated by calculating an upper and lower bound for each point in the surface and then marginalising these across one of the dimensions to obtain an uncertainty estimate about the mean 1D predicted posterior density.

In 2D and especially when considering sky localisation, we are also interested in the contours that enclose a given volume of probability density (usually the $50\%$ and $ 90\% $ contours), to plan optimal observation strategies in the search of electromagnetic counterparts. We propagate the uncertainty estimate produced by the GP (in the space of all realisations from the GP) to the physical parameter space to obtain the difference in the area enclosed by a given contour integral between an upper and lower bound on the density surface. To calculate the uncertainty for a given confidence level, e.g. $90\%$, we evaluate the upper and lower contour surfaces that correspond to the minimum probability density for the mean prediction contour which encloses $90\%$ of the volume.

Geometrically, we are building an uncertainty envelope around the mean GP prediction. The uncertainty on the contour is then given by the location in physical parameter space where the edges of that envelope intersect the plane defined by the mean prediction contour.

\begin{figure} 
    \includegraphics[scale=0.35]{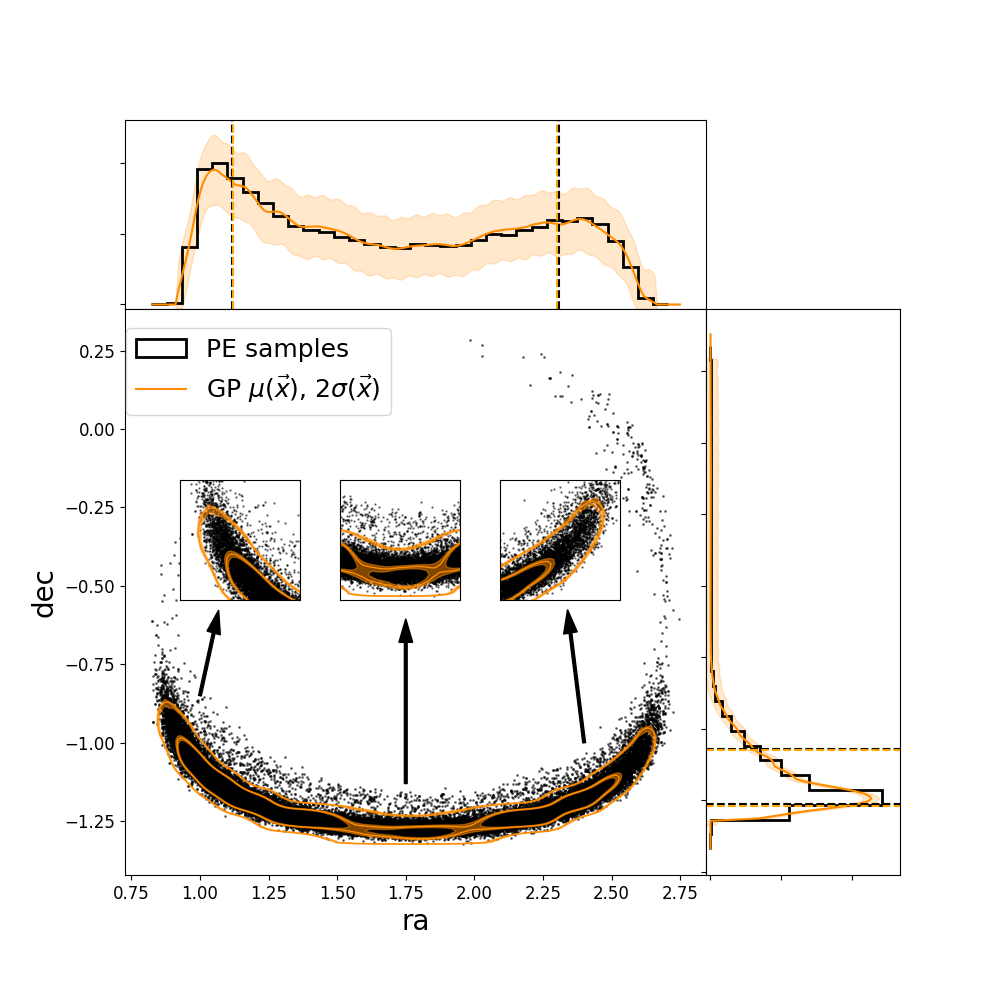}
    \caption{Central panel: contours of the 2D sky-location of GW150914, the GP model mean prediction and uncertainty (in orange) is compared to the points used to construct the fit (black crosses). Top and left panels show the GP model projections in 1D, compared to the original PE samples. All plots show the $2\sigma$ uncertainty around the density estimate as a shaded band}
    \label{fig:GP_2D}
\end{figure}

In the central panel of Fig~\ref{fig:GP_2D} we show the samples used to construct the model as well as the $50\%$  and $90\%$, contours of the GP interpolation in 2D with their respective $2\sigma$ uncertainty (the shaded regions). The top and left panels of Fig.~\ref{fig:GP_2D} show the mean prediction and its $ 2 \sigma$ uncertainty marginalised over each parameter by a simple integration of the density over its projection.

The inclusion of the uncertainty highlights several features. On the central inset in Fig~\ref{fig:GP_2D} we see that the lower bound on the $50\%$ contour is composed of three islands which correspond to peaks, while for both the mean and the upper bound these islands are connected to obtain a smooth surface at this contour level. For the outer $90\%$ contour we see that the differences mainly manifest in the tails, where as expected the upper bound follows the well known \emph{ring} around the sky slightly further. This matches our intuition that there is possibly more density around the ring than around the edges of the contour in the middle of the plot.

\section{Conclusions}\label{sec:conclusions}
We have presented an alternative method for density estimation of marginal PDFs for gravitational-wave parameters. Our method combines the desirable features of histograms to the extrapolation capabilities of KDEs, within a Bayesian framework. The choice of histogram binning determines the resolution of the PDF, while the kernel of the GP allows the interpolation to be flexible over non-Gaussian correlations and yet smooth. The noise variance parameter of the GP ensures that sources of stochastic noise from the histogram density estimation are taken into account. In cases where we employ an exact inference scheme, this noise variance can be evaluated for each histogram bin and it is equivalent to heteroskedastic errors over the density estimation. This allows to fully propagate the uncertainty from the PE samples. We plan to extend this method and fully incorporate uncertainties, as we showed in this work for the sky localisation example, over higher-dimensional posterior surfaces in future work. 

This method may be preferable to other methods such as KDEs, a closely related method which is sometimes adopted in the field, depending upon the use-case requirements. It comes with three main advantages: it is suitable for most interpolation problems commonly encountered for gravitational-wave marginal posteriors; it provides a Bayesian measure of uncertainty over the model predictions; it allows to quickly re-sample the interpolation using HMC and other samplers available in \tensorflow. 
We presented a series of examples where we know the accuracy of the interpolation is important, such as EOS calculations and sky localisation. As the number of events will increase in the next observing run $(O4)$, we need reliable tools to post-process the large volume of results. 

This work has highlighted the power of GPs to fit a gravitational-wave posterior surface, a natural extension of this work is to generate a surrogate for the entire likelihood surface, similar to what was done by the authors of~\citep{vivanco2019measuring} using a random forest regressor. Such use of GPs has been already investigated in the field of cosmology to model the Planck18 posterior distribution~\citep{McClintock2019}. This work has laid the foundation for us to apply a similar methodology to the gravitational-wave problem in a future work which is currently in preparation~\citep{demiliogwgplikelihood}. This has applications such as active sampling, efficient jump proposals~\citep{graff2012bambi, farr2020kombine} and more general use of the GP variance to guide the sampling process. The surface learned by the GP can be evaluated directly for a given set of parameters, therefore, avoiding the need to compute expensive waveforms. An example where such likelihood surrogates could be exploited is fast re-sampling with new astrophysical priors. This could replace an often difficult re-weighting procedure, especially when a prior assumption limits the number of available samples in a region of interest ~\citep{mandel2020GW190420}.


\section*{Acknowledgement}
We are grateful to Erik Bodin (Bristol University) and Dr. Carl Henrik-Ek (Cambridge University) for useful discussions. We thank Colm Talbot for their useful comments on the manuscript.
This work was supported by Science and Technology Facilities Council (STFC) grant ST/V001396/1, and we are grateful for the computational resources provided by Cardiff University and supported by STFC grant ST/V001337/1 (UK LIGO Operations award).
This research has made use of data, software, and/or web tools obtained from the Gravitational Wave Open Science Center (https://www.gw-openscience.org), a service of LIGO Laboratory, the LIGO Scientific Collaboration, and the Virgo Collaboration. LIGO is funded by the U.S. National Science Foundation. Virgo is funded by the French Centre National de Recherche Scientifique (CNRS), the Italian Istituto Nazionale Della Fisica Nucleare (INFN), and the Dutch Nikhef, with contributions by Polish and Hungarian institutes. Our code-base is developed upon \tensorflow~\cite{abadi2016tensorflow} and \gpflow~\cite{matthews2017gpflow}. Plots were prepared with \textit{Matplotlib}~\cite{hunter2007matplotlib} and the corner plots were made with Corner~\cite{foreman2016corner}.
\bibliographystyle{mnras}
\bibliography{references.bib}

\newpage
\begin{appendices}
\section{\label{apdx:techincal-gp} Technical details of the GP model}
\subsection{\label{apdx:data-preproc} Data pre-processing}
Data pre-processing, often referred to as data-set standardisation, is a common practice within the realm of machine learning and it can have a very high impact on the accuracy of the model. 
Our posterior samples have a wide range of values, some having bounds $[-1, 1]$ and some reaching $\mathcal{O}(10^3)$.
We re-scale our posterior samples such that each parameter ranges between $[0,1]$ by using the following transformation:
\begin{equation}
    \vec{\widetilde{\theta_{d}}} = \frac{(\vec{\theta_{d}} - min(\vec{\theta_{d}}))}{(max(\vec{\theta_{d}}) - min(\vec{\theta_{d}}))}
\end{equation}
where $\vec{\widetilde{\theta_d}}$ is the vector of transformed samples and the \textit{min} and \textit{max} are evaluated for each parameter (i.e. each dimension of the posterior samples vector).
The approximate marginalised posterior is scaled according to the \textit{z-score}, such that it has zero mean and unit variance:
\begin{equation}
     \widetilde{p}({\theta_{i}}|d) = \frac{p(\theta_{i}|d) - \mu_{p( \theta_{i}|d)}}{\sigma_{p(\theta_{i}|d)}}
\end{equation}
where $\widetilde{p}(\theta_{i}|d)$ is the transformed marginalised posterior, $\mu_{p(\theta_{i}|d)}$ and $\sigma_{p(\theta_{i}|d)}$ are respectively the mean and standard deviation of the marginalised posterior points.
All pre-processing in this work is performed using \texttt{Scikit-Learn}~\citep{scikit-learn}.

\subsection{\label{apdx:kernel-design} Kernel design}
The kernel is defined as the prior covariance between any two function values. Our prior knowledge about the morphology of the posterior can be encoded via this covariance, as it determines the space of functions that the GP sample paths live in.
The radial basis function (RBF) or squared exponential kernel is the most basic kernel and it's given as:
\begin{equation}
    \kappa_{\rm{RBF}}(x, x^{\prime}) = \sigma^2 \exp(-\frac{1(x - x^{\prime})^2}{2\ell^2})
\end{equation}
where the Euclidian distance between $(x, x^{\prime})$ is scaled by the length-scale parameter $\ell$ (measure of deviations between points) and the overall variance is denoted by $\sigma^2$ (average distance of the function away from its mean). Functions drawn from a GP with this kernel are infinitely differentiable.

For our application, a more complex kernel architecture that can capture the correlations between parameters is needed. We need smoothness over small scale features, such that we don't model random noise fluctuations of samples, and flexibility over the large scale characteristics of the posterior. For this purpose we employ a combination of RBF and
Matern, which is a generalisation of the RBF kernel with an additional smoothness parameter $\nu$. The smaller $\nu$, the less smooth the approximated function is:
\begin{multline}
    \kappa_{\rm{M\nu}}(x, x^{\prime}) = \sigma^2 \frac{2^{1-\nu}}{\Gamma(\nu)}\left(\sqrt{2\nu}\frac{(x-x^{\prime})}{\ell}\right)^{\nu}\\
    K_{\nu}\left(\sqrt{2\nu}\frac{(x-x^{\prime})}{\ell}\right)
\end{multline}
We choose $\nu=(\frac{1}{2},\frac{5}{2})$ depending on the specific morphology of the posterior, as this kernel is responsible for encoding its overall shape such as sharp boundary features. 
The resulting kernel equation is given by:
\begin{equation*}
    \kappa_{GP}(\vec{\theta_{d}}, \vec{\theta_{d}}^{\prime}) = \kappa_{\rm{RBF}} \times \kappa_{\rm{M52}}
\end{equation*}
The kernel multiplication is equivalent to an AND operation, as it corresponds to an element-wise multiplication of their corresponding covariance matrices. This means that the resulting covariance matrix will only have a high value if both covariances have a high value.
We also apply automatic relevance determination (ARD), which modifies the kernel such that for each dimension an appropriate length scale is chosen \citep{neal2012bayesian}. 
 \label{apdx:techincal-gp}
\end{appendices}

\end{document}